\documentclass{iaus}
\usepackage{graphicx}

\title[High-gravity central stars]{High-gravity central stars}

\author[Thomas Rauch]{Thomas Rauch}%\thanks{}

\affiliation{Institut f\"ur Astronomie und Astrophysik, Universit\"at T\"ubingen, Germany\break 
             email: rauch@astro.uni-tuebingen.de}

\pubyear{2006}
\volume{234}  %% insert here IAU Symposium No.
\jname{Planetary Nebulae in our Galaxy and Beyond}
\editors{M. J. Barlow \& R. H. M\'endez, eds.}

\begin{document}

\maketitle

\begin{abstract}
NLTE spectral analyses of high-gravity central stars by means of state-of-the-art
model atmosphere techniques provide information about the precursor AGB stars. 
The hydrogen-deficient post-AGB stars allow investigations on the intershell matter 
which is apparently exhibited at the stellar surface.
We summarize recent results from imaging, spectroscopy, and spectropolarimetry.

\end{abstract}

\section{Introduction}
High-gravity central stars (CS) of planetary nebulae (PNe) display the hottest stage of 
stellar post-AGB evolution just at the entrance of the white-dwarf cooling track on the 
way towards the stellar graveyard.
These stars are in a short phase between decreasing mass loss and thus, less spectral 
contamination by features which are formed under the influence of strong stellar winds 
(cf\@. Kudritzki, these proceedings), and increasing gravity which, in interplay with 
radiative levitation, will wipe out all photospheric information about their progenitors. 

We will show progress in analyses of high-gravity CSPNe and demonstrate that
spectral analyses by means of non-LTE model-atmosphere techniques (Sect\@. 
\ref{sect:spectralanalyses}) provide a powerful tool to investigate on properties of 
these stars and to determine constraints for evolutionary theory (Sect\@. \ref{sect:pg1159}).

\section{Imaging and spectroscopy}
\label{sect:spectralanalyses}

High-gravity post-AGB stars, i.e\@. hot, compact stars, have arrived at extremely high
effective temperatures ($T_\mathrm{eff}$ up to $\approx 200\,\mathrm{kK}$). Thus it appears 
that their flux maximum is located in the EUV spectral range and they are intrinsically
faint in the optical. For a reliable spectral analysis, the evaluation of ionization equilibria
is necessary for the determination of $T_\mathrm{eff}$ (see Sect.\,\ref{sect:lsv4621}). 
Since the strongest metal lines are found almost exclusively in the high-energy range, its accuracy 
is dependent on the availability of high-resolution and high-S/N spectra in the UV wavelength range. 
These have been provided e.g\@. by the Hubble Space Telescope (HST) with
the Faint Object Spectrograph (FOS) and the Space Telescope Imaging Spectrograph (STIS),
and the Far Ultraviolet Spectroscopic Explorer (FUSE) \ref{tab:sat}). 
Examples for analyses which have been performed based
on data obtained with these (\ref{tab:sat}) are given in Sections \ref{sect:xMC} and \ref{sect:pg1159}.

\begin{table}
  \begin{center}
  \caption{Satellite instruments used for spectral analyses of high-gravity CS.}
  \label{tab:sat}
  \begin{tabular}{llr@{-}lr@{-}rr}\hline
%    \noalign{\smallskip}
    satellite & instrument & \multicolumn{2}{c}{period} & \multicolumn{2}{c}{$\lambda / \mathrm{\AA}$} & resolution \\
    \hline
    HST        & FOS        & 1990 & 1997 & 1150 & 8000 &        $< 1\,300$ \\
    HST        & STIS       & 1997 & 2004 & 1150 & 3175 &      $< 114\,000$ \\
    FUSE       &            & 1999 &      &  905 & 1185 & $\approx 20\,000$ \\
    \hline
  \end{tabular}
 \end{center}
\end{table}

\subsection{LMC + SMC central stars}
\label{sect:xMC}

In order to constrain the initial-to-final mass relation, analyses of 
35 LMC \cite{VEA03} and 
27 SMC \cite{VEA04} CS have be performed, 
based on imaging and spectroscopy by HST STIS and the Wide-Field Planetary Camera (WFPC2).
A comparison in the Hertzsprung-Russell diagram with evolutionary tracks of \cite{VW94} for LMC and SMC
metallicities yields average CS masses of 0.65\,M$_\odot$ and 0.63\,M$_\odot$, respectively.
Although these analyses are free of distance uncertainty, the determination of $T_\mathrm{eff}$
by the Zanstra method imposes an imponderable error which one has to be aware of.

An analysis of seven LMC CS has been presented by \cite{HB04}. They used FUSE and
HST (FOS + STIS) spectra and employed NLTE model-atmosphere techniques using the codes
TLUSTY \cite{HL95} and, since most objects display wind features 
($\dot{M} \approx 5\times 10^{-8}\,\mathrm{M_\odot yr^{-1}}$), CMFGEN \cite{HM98, HM99}.
The CS's $T_\mathrm{eff}$  range from 38 to 70\,kK and five of them have LMC abundances. 
Since the surface gravity $g$ could not be determined from the UV spectra and adequate optical
spectra are not available, it is adopted within $4.3 \leq \log g \leq 5.3\,\mathrm(cgs)$ in 
order to investigate on errors in the parameter determination. However, this analysis represents 
not the standard of modern spectral analysis because $g$ is not self-consistently determined. 
One has to be aware of this uncertainty although this analysis aims on a characterization of PNe by 
better understanding of the CS $\leftrightarrow$ PN system and on wind properties.

\begin{figure}\centering
 \includegraphics{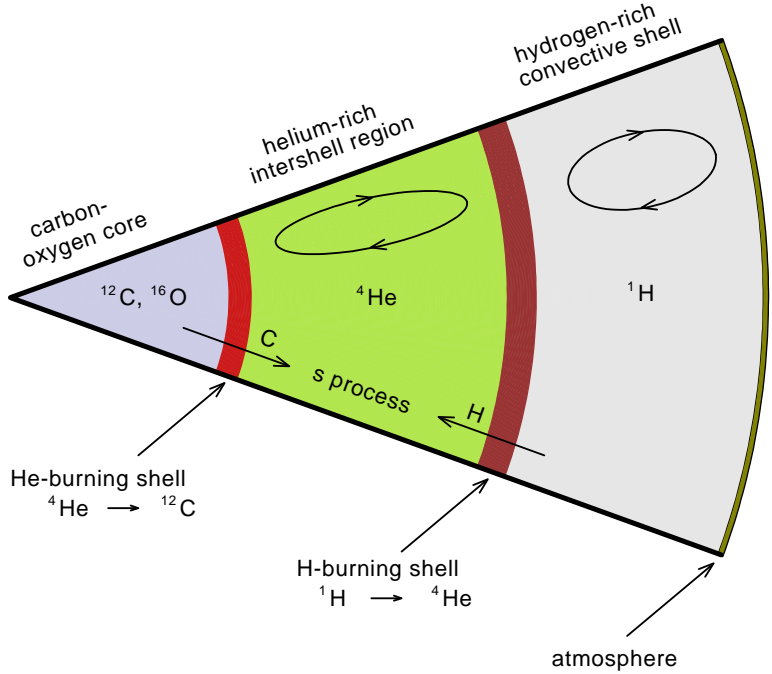}
  \caption{Interior of an AGB star. Between the He- and H-burning shells, the so-called intershell
           matter is located ($\approx 10^{-2}\,\mathrm{M_\odot}$). The H-rich envelope
           ($\approx 10^{-4}\,\mathrm{M_\odot}$) prohibits an insight into the intershell matter.
           A (V)LTP may lay bare this matter.}
  \label{fig:interior}
\end{figure}

\subsection{Hydrogen-rich central stars}
\label{sect:normal}

Newly available high-resolution UV spectra, obtained by HST STIS and FUSE, enabled \cite{TEA05} 
to establish an improved temperature scale for seven hot hydrogen-rich CSPN. Within their
NLTE model-atmosphere analysis, they evaluated ionization balances of C, N, and O which show 
rich line spectra in the UV in order to derive $T_\mathrm{eff}$ precisely. 
\cite{HoEA05} extended this analysis to the use of the Fe\,{\sc vi}\,/\,Fe\,{\sc vii} ionization 
equilibrium.

In an on-going analysis of LS\,V+4621 \cite{ZEA06}, all iron-group elements are considered individually.
In Sect\@. \ref{sect:lsv4621}, we show some details of this state-of-the-art spectral analysis.

\subsection{Hydrogen-deficient central stars}
\label{sect:pg1159}

While the hydrogen-rich stellar post-AGB evolution is quite well understood, the picture of 
hydrogen-deficient evolution (about 20\% of all post-AGB stars) became 
clearer only in the last years (cf\@. Herwig, these proceedings).
Spectral analysis of these objects has placed challenges again and again to evolutionary theory and, vice versa,
predictions from evolutionary model have been verified by spectral analysis.

One example is the discovery of the Ne\,{\sc vii} $\lambda 3643.6\,\mathrm{\AA}$ absorption line
in optical spectra of three extremely hot PG\,1159-type CSPN \cite{WR94}. 
An abundance analysis with NLTE model atmospheres has shown that
this line can be reproduced with a Ne abundance of $\approx 2\,\mathrm{\%}$ (by mass) in the photosphere.
Early calculations of \cite{IT85} could not explain the observation, although they predicted Ne to be the
fourth abundant element ($\approx 2\,\mathrm{\%}$ mass fraction, following C, O, and He) just below the 
helium-burning shell -- inaccessible for spectroscopy. 
New evolutionary models by \cite{HEA99}, which consider a realistic overshoot adopted from 
hydrodynamical simulations, predict a mass fraction of about 3.5\,\% of Ne in the intershell
(Fig\@. \ref{fig:interior}), in good agreement with the spectroscopic result. 
The recent identification of the Ne\,{\sc vii} $\lambda 973.3\,\mathrm{\AA}$ absorption line in 
FUV \cite{WEA04b} spectra (obtained with FUSE) of six PG\,1159 stars,
the presence of Ne\,{\sc vii} $\lambda 3643.6\,\mathrm{\AA}$ in eight and
of Ne\,{\sc vii} $\lambda\lambda 3850-3910\,\mathrm{\AA}$ in four PG\,1159 stars \cite{WEA04b} 
corroborates these evolutionary models. Stellar atmospheres with 
a content of 2\,\% of Ne (eleven times solar) can reproduce well the observed Ne\,{\sc vii} lines. 
It its worthwhile to note that \cite{HEA05} have shown that the Ne\,{\sc vii} $\lambda 973\,\mathrm{\AA}$ 
line appears with a strong P Cygni wind profile in the CS of the PNe NGC 2371, Abell 78, and K\,1-16 
(this is much stronger than the supposed C\,{\sc iii} $\lambda 977\,\mathrm{\AA}$ P Cygni feature in these CS). 

Another example is the reproduction of the CHANDRA spectrum of the 
hydrogen- and helium-deficient star H\,1504+65 \cite{WEA04a}. The consideration of magnesium with a mass fraction of 2\%
significantly improved the spectral fit. For Mg \cite{IT85} predicted an equal amount like that of Ne,
also just below the helium-burning shell.

The reason for our special interest in hydrogen-deficient post-AGB stars, especially in the PG\,1159 stars,
lies in the fact, that these experience a (very) late thermal pulse (cf\@. Herwig, these proceedings) 
during their evolution which mixes the entire
hydrogen-rich envelope (Fig.\,\ref{fig:interior}) into the intershell matter where hydrogen is burned then.
Spectroscopy of these stars provides a direct view on the former intershell matter and allows to conclude
on details in nuclear and mixing processes in AGB stars.

We will now briefly summarize PG\,1159 abundance determinations based on recent HST and FUSE spectrocopy.
He, C, N, O, Ne, Mg, F (Sect.\,\ref{sect:fluorine}), and Si are in line with predictions from evolutionary models
\cite{ReEA05} and these proceedings, and Jahn et al\@. 2006).
Fe (Sect.\,\ref{sect:iron}) shows a surprisingly large depletion, as well as S (Sect.\,\ref{sect:sulfur}).
P is predicted to be overabundant by a factor of up to 25 (still uncertain) but is found to be roughly
solar \cite{JEA06}.

For a more detailed review on elemental abundances in bare CSPN, see \cite{WH06} and references therein.

\subsubsection{Iron}
\label{sect:iron} 

Evolutionary calculations predict a reduced, i.e\@. sub-solar, Fe intershell abundance due to n-captures in
the s-process. The FUSE wavelength range covers the strongest Fe\,{\sc vii} lines (this is the dominant
ionization stage in hot PG\,1159 stars). Up to now, FUSE spectra of three PG\,1159 stars with sufficiently 
high S/N have been analyzed but no iron lines are detectable. Thus, a very strong Fe depletion (at least
$1 - 2\,\mathrm{dex}$) takes place in the intershell \cite{MEA02}.

\subsubsection{Fluorine}
\label{sect:fluorine} 

The first discovery of fluorine (F\,{\sc vi}\,$\lambda 1139.50\,\mathrm{\AA}$) in FUSE spectra of hot post-AGB stars
\cite{WEA05} has shown that AGB stars are efficiently synthesizing fluorine which even survives a (V)LTP -- 
abundances of up to $200\times$ solar have been determined in PG\,1159 stars. Measurements in PNe (Zhang \& Liu,
these proceedings) have found that F is generally overabundant in PNe, thus providing new evidence for its
synthesis in AGB stars.

\subsubsection{Sulfur}
\label{sect:sulfur} 

\begin{figure}
 \includegraphics{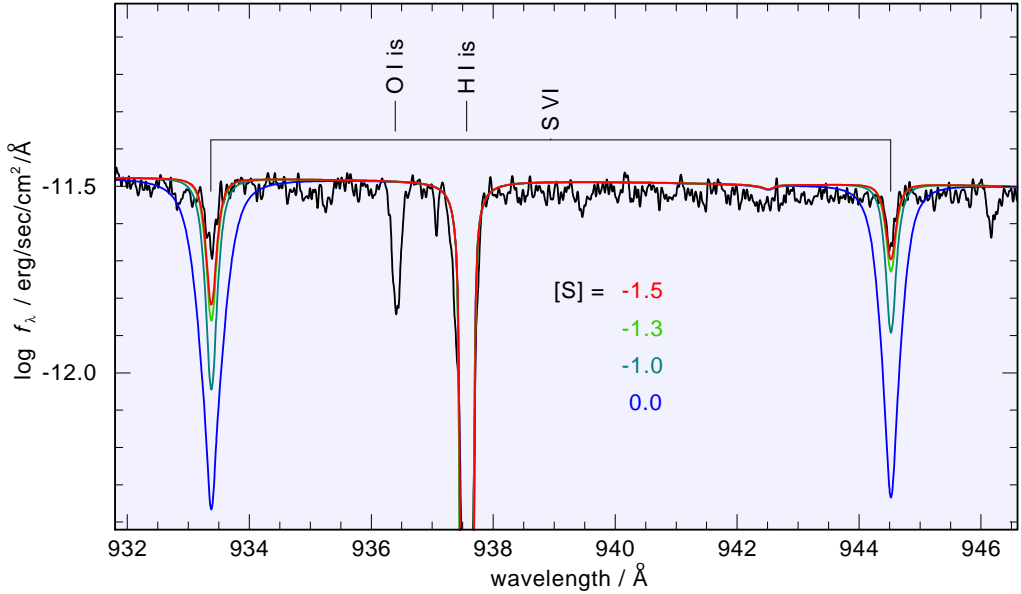}
  \caption{Section of the STIS spectrum of PG\,1159$-$035 showing the S\,{\sc vi} resonance doublet. Obviously,
           S is strongly underabundant (about 0.02 of the solar abundance).}
  \label{fig:sulfur}
\end{figure}

The photospheric abundance of sulfur should not be changed by nuclear processes. A recent spectral
analysis of PG\,1159$-$035 \cite{JEA06} has shown, however,  that S is strongly underabundant (Fig\@. \ref{fig:sulfur}). 

\section{Spectropolarimetry}
\label{sect:magnetic}

Recently, \cite{JEA05} reported on the first directly measured (spectropolarimetry with 
the Focal Reducer and low dispersion Spectrograph (FORS1) attached to the 
Very Large Telescope (VLT))
magnetic fields ($B$ of the order of kG) in four CS (of the PNe NGC\,1360, EGB\,5, LSS\,1362, A\,36).
From these field strengths (under assumption of complete flux conservation), they estimate that the 
main-sequence precursors had field strengths of the order of some G. The later white dwarfs (WDs) 
will have fields of some MG. Many WDs with such strong fields are indeed known.
However, the influence of magnetic fields on the morphology of PNe is still under debate and this first
measurement is a valuable help to quantify their impact in magneto-hydrodynamical modeling.

A possible measurement of the Zeeman splitting (cf\@. Stanghellini et al\@., these proceedings) of an O\,{\sc IV} 
line in a high-resolution optical spectrum of the CSPN of He\,2-36 yields even a higher field strength of 
$B \approx 40\,\mathrm{kG}$.

\section{Imaging}
\label{sect:PNsearch}

PNe are detected only around every other PG\,1159 star. Depending on their individual evolution, the PNe which have
been ejected on the AGB have most likely dispersed under the detection limit. Over the last years we spent
some time with the search of nebular emission around PG\,1159 stars and related objects.
Our search has been entirely negative but for DO white dwarf PG\,0109+111 \cite{WEA97} where a possible PN was found with
a diameter of 0.8\,pc. Recently, \cite{HEA03} discovered a PN (Hewett 1) around the DO PG\,1034+001 with a diameter of
$3.5 - 7.0\,\mathrm{pc}$. An investigation on data of the Southern H-Alpha Sky Survey Atlas (SHASSA, \cite{GEA01})
by \cite{REA04} has shown that an enormous ionized
halo exists around this object with a diameter of $16 - 24\,\mathrm{pc}$. An ionized nebula has also been
found by \cite{CEA04} around the DO KPD\,0005+5106. Since the ionized mass is about $70\,\mathrm{M_\odot}$ this is
unambiguously ambient interstellar medium (ISM). \cite{CEA04} also suggested that the nebula around PG\,1034+001 is
simply ionized ISM. Frey (these proceedings) also expects this for PG\,0109+111's nebula. Although all these emissivities
are probably no PNe, the importance of hot white dwarfs as ionizing sources for the ISM is quite evident.

\section{Evolutionary Theory}
\label{sect:evolution}

Recently, new calculations for post-AGB evolution have been presented by \cite{AEA05a} (hydrogen-deficient non-DA white
dwarfs), \cite{AEA05b} (DA white dwarfs), \cite{MBEA06} (born-again scenario), and \cite{MA06} (PG\,1159 and O(He) stars).
These new calculations should be used for, e.g., the determination of white-dwarf masses (Herwig, priv\@. comm.). 
\cite{MA06} find a systematic offset of $\Delta M \approx -0.05\,\mathrm{M_\odot}$ for the known PG\,1159 stars 
compared to the values summarized in \cite{WH06} which had been determined from older evolutionary calculations.

\section{FUSE and HST spectroscopy of LS\,V+4621}
\label{sect:lsv4621}

LS\,V+4621 is the exciting star of Sh\,2-216. This is the closest ($d=130\,\mathrm{pc}$) PN known, with an
apparent diameter of 1.6 degrees projected at the sky. High-resolution FUV and UV spectra
(FUSE: 67.6 ksec in 2003/2004, $\Delta\lambda\approx 0.05\,\mathrm{\AA}$, 
STIS: 5.5 ksec in 2000, $\Delta\lambda\approx 0.06\,\mathrm{\AA}$)
are available. These have been analyzed by means of NLTE model-atmosphere techniques. We employed
{\sc TMAP}, the T\"ubingen NLTE Model Atmosphere Package \cite{WEA03} in order to
calculate models which consider H+He+C+N+O+Mg+Si by ``classical'' model atoms and
Ca-Ni in a statistical approach \cite{RD03} introducing superlevels and superlines. In total, 531 levels are treated
in NLTE, with 1761 individual lines and about nine million iron-group lines \cite{K96}.

The HST STIS spectrum allows to determine the interstellar neutral hydrogen density
$N_\mathrm{H\,{\sc I}} = 1.10\pm 0.05\cdot 10^{20}\,\mathrm{cm^{-2}}$ and the interstellar extinction
$E_\mathrm{B-V} = 0.03$. The evaluation of ionization equilibria of
N\,{\sc iv}\,/\,N\,{\sc v},
O\,{\sc iv}\,/\,O\,{\sc v},
Si\,{\sc iv}\,/\,Si\,{\sc v},
Fe\,{\sc v}\,/\,Fe\,{\sc vi}\,/\,Fe\,{\sc vii}, and
Ni\,{\sc v}\,/\,Ni\,{\sc vi}
allow the determination of $T_\mathrm{eff} = 95\pm 3\,\mathrm{kK}$ and $\log g = 6.9\pm 0.2\,\mathrm{(cgs)}$
with unprecedented precision.

\begin{figure}
 \includegraphics{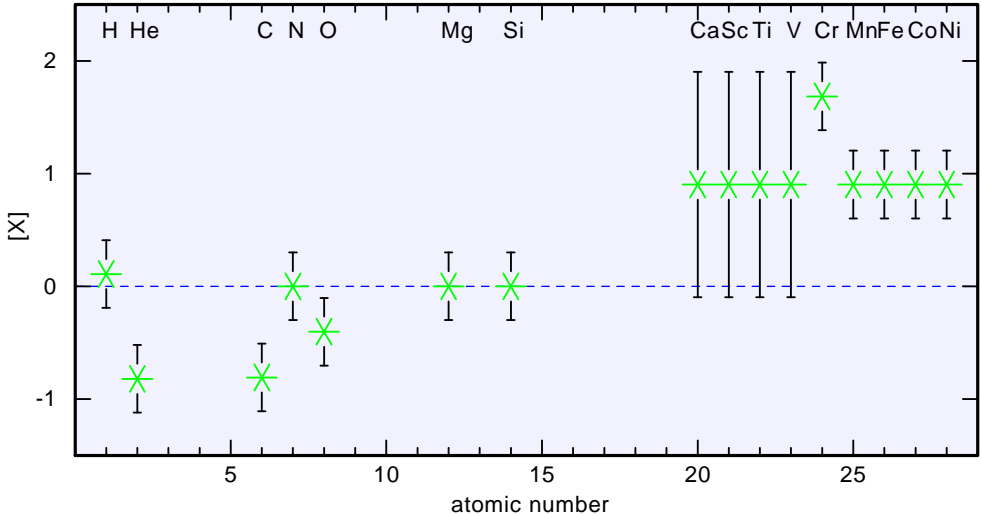}
  \caption{Photospheric abundances in LS\,V+4621. [X] denotes log (abundance / solar abundance).
           The abundance pattern indicates that gravitational settling is already efficient for He 
           while radiative levitation causes strong overabundances of the iron-group
           elements. The large error ranges for Ca, Sc, Ti, and V are due to the fact that no
           Kurucz POS lines can be identified in the STIS spectrum. Since the FUSE spectrum is
           strongly contaminated by interstellar absorption, the abundance determination for
           these elements is still less certain.}
  \label{fig:abundances}
\end{figure}

\begin{figure}
 \includegraphics{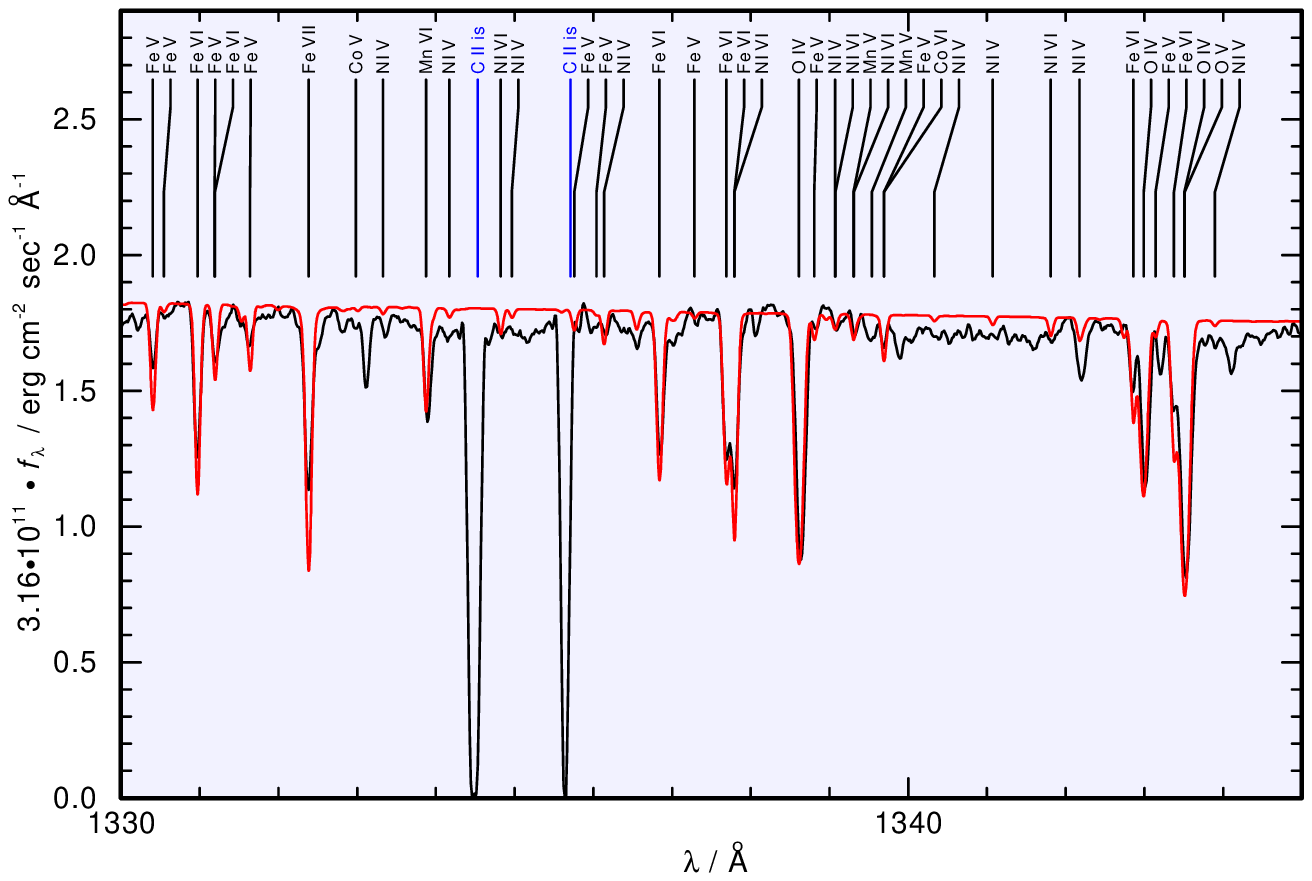}
  \caption{Section of the HST STIS spectrum of LS\,V+4621 (shifted by $v_\mathrm{rad} = 20.386\,\mathrm{km/sec}$
           in order to match the rest wavelengths of the photospheric lines)
           compared with the theoretical spectrum ($T_\mathrm{eff} = 95\,\mathrm{kK}$, $\log g = 6.9$,
           abundances see Fig\@. \ref{fig:abundances}).
           The identified lines are marked. is denotes lines of interstellar origin. 
           Note that e.g\@. lines from Fe\,{\sc v}, Fe\,{\sc vi}, and Fe\,{\sc vii} are matched simultaneously. 
           For the calculation of the synthetic spectrum,
           Kurucz's POS line lists \cite{K96}, which contain lines with laboratory measured wavelengths only, 
           are used. For the model-atmosphere calculation Kurucz's LIN line lists are used (including 
           all theoretically calculated lines in addition). This may explain the existing differences between observation
           and model, e.g\@. unidentified features.}
  \label{fig:obsmod}
\end{figure}

Si\,{\sc v} has been newly identified.
Mg\,{\sc iv} lines have been discovered for the first time in these objects \cite{ZEA06}.
The determined abundances are summarized in Fig\@.\,\ref{fig:abundances}. A comparison
of the synthetic model with the observation is shown in Fig\@. \ref{fig:obsmod}.

\section{NLTE model-atmosphere fluxes}
\label{sect:fluxes}

Recent progress in observational techniques as well as numerical methods facilitate to examine 
closely both, PNe as well as their central stars. NLTE model-atmosphere calculations have arrived 
at a high level of sophistication, providing reliable ionizing fluxes to be used to construct
adequate photoionization models of PNe. The use of a black-body approximation instead appears
attractive because it is so-easy to use. However, a challenge for the next years is to improve 
PNe modeling with a reliable consideration of the central stars' fluxes.

Grids of flux tables calculated for high-gravity CS and variety of photospheric chemical compositions
can be retrieved from {\tt http://astro.uni-tuebingen.de/\raisebox{1mm}{\tiny $\sim$}rauch}.
These can be used as ionizing fluxes in photoionization codes such as
{\sc CLOUDY} \cite{F03} or  
{\sc MOCASSIN} (cf\@. Ercolano these proceedings). 
In case of any special requirements please do not hesitate to ask the author for an individual model.

\begin{acknowledgments}
I would like to thank Klaus Werner for careful reading of the manuscript.
This research was supported by the DLR under grant 50\,OR\,0201 and by a grant
of the DFG (KON 218/2006).
This research has made use of NASA's Astrophysics Data System and of 
the SIMBAD Astronomical Database, operated at CDS, Strasbourg, France.
\end{acknowledgments}

\end{document}